\documentstyle[epsf,graphics,a4,12pt]{article}
\begin{document}
\newcommand{\beq}{\begin{equation}}
\newcommand{\eeq}{\end{equation}}
\newcommand{\beqn}{\begin{eqnarray}}
\newcommand{\eeqn}{\end{eqnarray}}
\newcommand{\dpf}{\displaystyle\frac}
\newcommand{\no}{\nonumber}
\newcommand{\ep}{\epsilon}
\begin{center}
{\Large Geometry and topology of two kinds of extreme
Reissner-Nordstr$\ddot{o}$m-anti-de Sitter black holes}
\end{center}
\vspace{1ex}
\centerline{\large Bin
Wang$^{a,b,}$\footnote[1]{e-mail:binwang@fma.if.usp.br},
\ Elcio Abdalla$^{a,}$\footnote[2]{e-mail:eabdalla@fma.if.usp.br} and
Ru-Keng Su$^{c,}$\footnote[3]{e-mail:rksu@fudan.ac.cn}}
\begin{center}
{$^{a}$ Instituto De Fisica, Universidade De Sao Paulo, C.P.66.318, CEP
05315-970, Sao Paulo, Brazil \\
$^{b}$ Department of Physics, Shanghai Teachers' University, P. R. China
\\
$^{c}$ Department of Physics, Fudan University, Shanghai 200433, P. R.
China}
\end{center}
\vspace{6ex}
\begin{abstract}
Different geometrical and topological properties have been shown for two
kinds of extreme Reissner-Nordstr$\ddot{o}$m-anti-de Sitter black holes.
The relationship between the geometrical properties and the intrinsic
thermodynamical properties has been made explicit.
\end{abstract}
\vspace{6ex} \hspace*{0mm} PACS number(s): 04.70.Dy, 04.20.Gz, 04.62.+v.
\vfill
\newpage
The study of the extreme black hole (EBH) has been stimulated since the
discovery [1,2] that the four-dimensional (4D) Reissner-Nordstr$\ddot{o}$m
(RN) EBH is a different object from its nonextreme counterpart owing to
its drastically different topological properties and peculiar zero entropy
regardless of its nonzero horizon area. However, these results met some
challenges subsequently. Starting with the grand canonical ensemble, it
was argued [3-5] that in a finite size cavity a 4D RN non-extreme black
hole (NEBH) can approach the extreme state as closely as one likes and the
geometrical and topological properies are still of nonextreme sectors.
Bekenstein-Hawking formula is believed to hold for RN EBH entropy
description, which is also supported by state-counting calculations of
certain extreme and near-extreme black holes in string theory, see [6] for
a review. These different results indicate that EBHs have a controversial
role in black hole thermodynamics and topologies and require special care.

Comparing [1,2] and [3-5], it seems that the clash comes from two
different treatments: one refers to Hawking's treatment by starting with
the original EBH [1,2] and the other Zaslavskii's treatment by first
taking the boundary limit and then the extreme limit to get the EBH from
its nonextreme counterpart [3-5]. Applying these two treatments, it was
found that two different topological objects represented by different
Euler characteristics exist for 4D RN EBH, charged dilaton EBH [7], Kerr
EBH [8] as well as two-dimensional (2D) EBHs [7,9]. Drastically different
intrinsic thermodynamical properties have also been displayed for 4D Kerr
EBH [8] and 2D EBHs [9,10] due to these different treatments.
Based upon these results it was
suggested that there maybe two kinds of EBHs in nature: the first kind
suggested by Hawking et al with the extreme topology and zero entropy,
which can only be formed by pair creation in the early universe, while
the second kind, suggested by Zaslavskii, has the topology of
the nonextreme sector and the entropy is still described by the
Bekenstein-Hawking formula, which can be developed from its nonextreme
counterpart through second order phase transition [11-13]. This
speculation has been further confirmed recently in a Hamiltonian framework
[14] and the grand canonical ensemble [15] as well as canonical ensemble
[16] formulation for RN anti-de Sitter (AdS) black hole, where the
Bekenstein-Hawking entropy and zero entropy have been found again for RN
AdS EBHs. However the clear pictures of geometry and topology for two
kinds of RN AdS EBHs have not been presented therein.

The study of RN AdS black holes is appealing. This is not only
because it is a standard example to study the AdS/CFT correspondence
[17], but also because some striking resemblance of the RN AdS phase
structure to that of the Van de Waals-Maxwell liquid-gas system has been
observed and some classical critical phenomena has also been uncovered
[18,19]. It is of interest to investigate the geometrical and topological
properties of RN AdS EBHs and their relation to the EBHs' intrinsic
thermodynamics. We hope that detailed understanding of geometrical and
topological properties in RN AdS EBHs will help us to get a clear picture
of
the phase transition in RN AdS black holes.
This is the motivation of the pesent paper.

The RN black hole solution of Einstein's equations in free space with a
negative cosmological constant $\Lambda = - 3/l^2$ is given by
\beq    
{\rm d}s^2=-h{\rm d}t^2+h^{-1}{\rm d}r^2+r^2{\rm d}\Omega^2, A=Q/r{\rm
d}t,
\eeq
with
\beq    
h=1-\dpf{r_+}{r}-\dpf{r_+^3}{l^2 r}-\dpf{Q^2}{r_+
r}+\dpf{Q^2}{r^2}+\dpf{r^2}{l^2}.
\eeq
The asymptotic form of this spacetime is AdS. There is an outer horizon
located at $r=r_+$. The mass of the black hole is 
\beq          
M=\dpf{1}{2}(r_+ +\dpf{r_+^3}{l^2}+\dpf{Q^2}{r_+}).
\eeq
The Hawking temperature is given by the expression
\beq  
T_H =\dpf{1-\dpf{Q^2}{r_+^2}+\dpf{3r_+^2}{l^2}}{4\pi r_+}
\eeq
and the potential by 
\beq 
\phi =\dpf{Q}{r_+}
\eeq
In the extreme case  $r_+, Q$ satisfy the relation
\beq   
1-\dpf{Q^2}{r_+^2}+\dpf{3r_+^2}{l^2}=0.
\eeq 

Making use of the approach developed in [20], for a grand canonical
ensemble the RN
AdS black hole is considered in a cavity with radius $r_B$, and the local
temperature on the cavity is 
\beq     
T=\dpf{T_H}{[h(r_B)]^{1/2}}=\dpf{T_H}{[(1-\dpf{Q^2}{r_+
r_B}+\dpf{3r_B^2}{l^2})-\dpf{r_+}{r_B}(1-\dpf{Q^2}{r_B
r_+}+\dpf{r_+^2}{l^2}+\dpf{2r_B^3}{l^2 r_+})]^{1/2}}.
\eeq
When the black hole approaches the extreme state, $T_H\rightarrow 0$, the
simplest choice is to take $T\rightarrow 0$. One might refer to the third
law of thermodynamics to argue that the EBH cannot be achieved.

However, it is interesting to point out that although $T_H\rightarrow 0$,
the square root in (7) tends to zero as well if we take $r_+\rightarrow
r_B$. Thus the extreme state of RN AdS black hole with nonzero local
temperature can be achieved with no contradiction with the third law. In
the grand canonical ensemble actually only the temperature on the boundary
has physical meaning, whereas $T_H$ can always be rescaled without
changing observable quantities [21].

Now it is of interest to study the proper distance between a horizon and
any fixed point. Different behaviors of proper distances obtained in [1]
and [3] for RN EBHs are cognitive roots to qualitatively different two
types of EBHs. In our metric (1), this is the quantity
\beq  
L=\int
_{r_+}^{r_B}\dpf{dr}{\sqrt{h(r)}}=\int_{r_+}^{r_B}\dpf{rdr}
{\sqrt{\dpf{r^4}{l^2}+r^2-(r_+ +\dpf{r_+^3}{l^2}+\dpf{Q^2}{r_+})r+Q^2}} 
\eeq
To study the behavior of $L$ for the RN AdS EBH, we have two limits
to deal with, one is the extreme limit and the other is the boundary
limit. To perform the limit procedures,  we may take
$Q^2=(1+\dpf{3r_+^2}{l^2})r_+^2+\ep, \ep\rightarrow 0^+$ and $r_B=r_+
+\eta, \eta\rightarrow 0^+$, where $\ep, \eta$ are infinitesimal
quantities with different orders of magnitude. For Hawking's treatment,
starting with the original RN AdS EBH, the proper distance can be written
as 
\beq   
L_1=\int_{r_+}^{r_B}\dpf{rdr}{\sqrt{(r-r_+)^2+(r^4-4r_+^3 r+3r_+^4)/l^2}}
\eeq
By taking the boundary limit $r_+\rightarrow r_B$, this quantity diverges
(Fig.1). However for Zaslavskii's treatment by first taking the boundary
limit and then the extreme limit, we can evaluate the behavior of the
proper distance $L_2$ mathematically by taking $\eta=\ep^p, p>1$, it leads
to a finite value as shown in Fig.2. These results are consistent with
those obtained in RN cases [1,3].

\begin{figure}[htb]
\begin{center}
\leavevmode
\begin{eqnarray}
\epsfxsize= 5truecm\rotatebox{-90}{\epsfbox{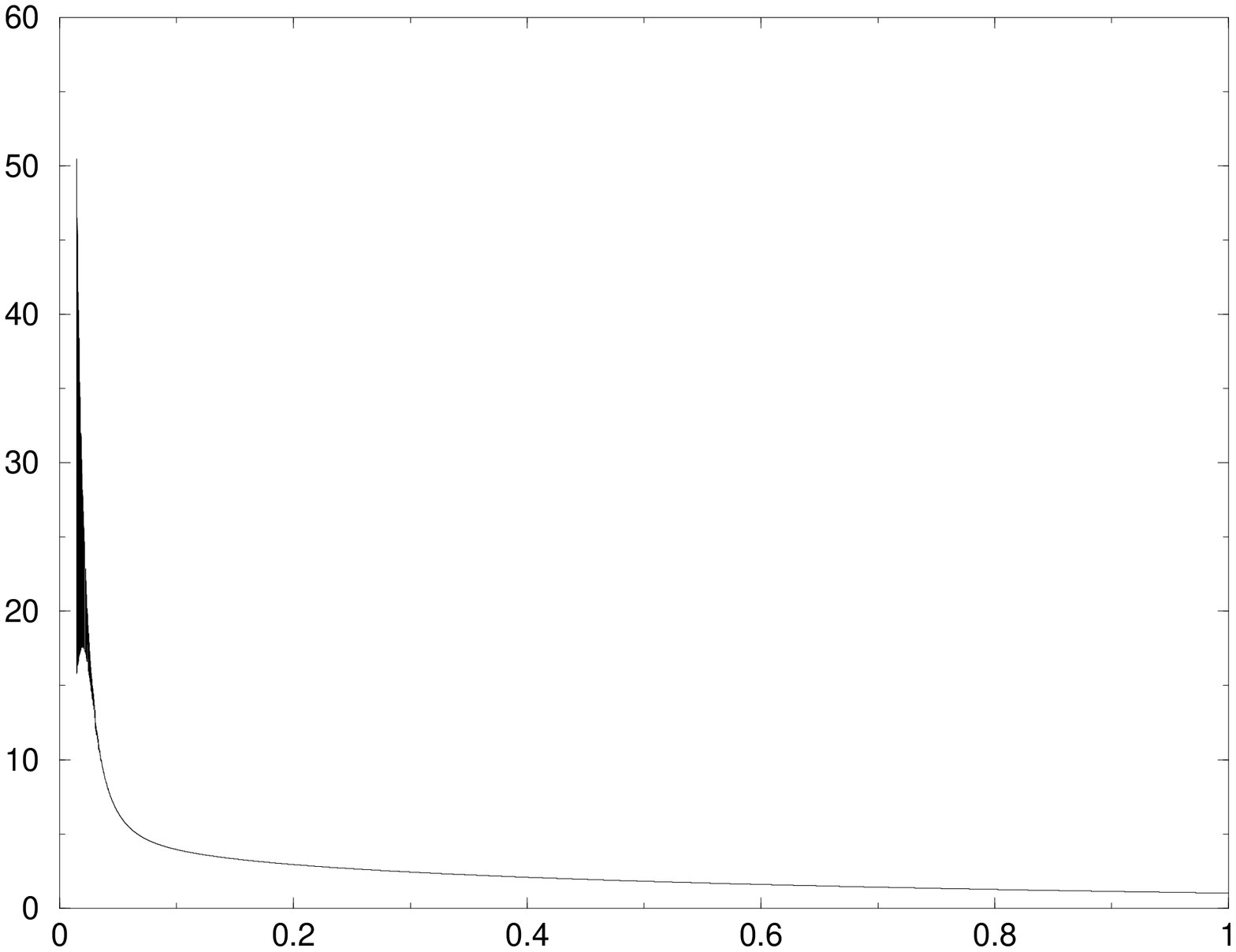}} & &
\epsfxsize=5truecm\rotatebox{-90}{\epsfbox{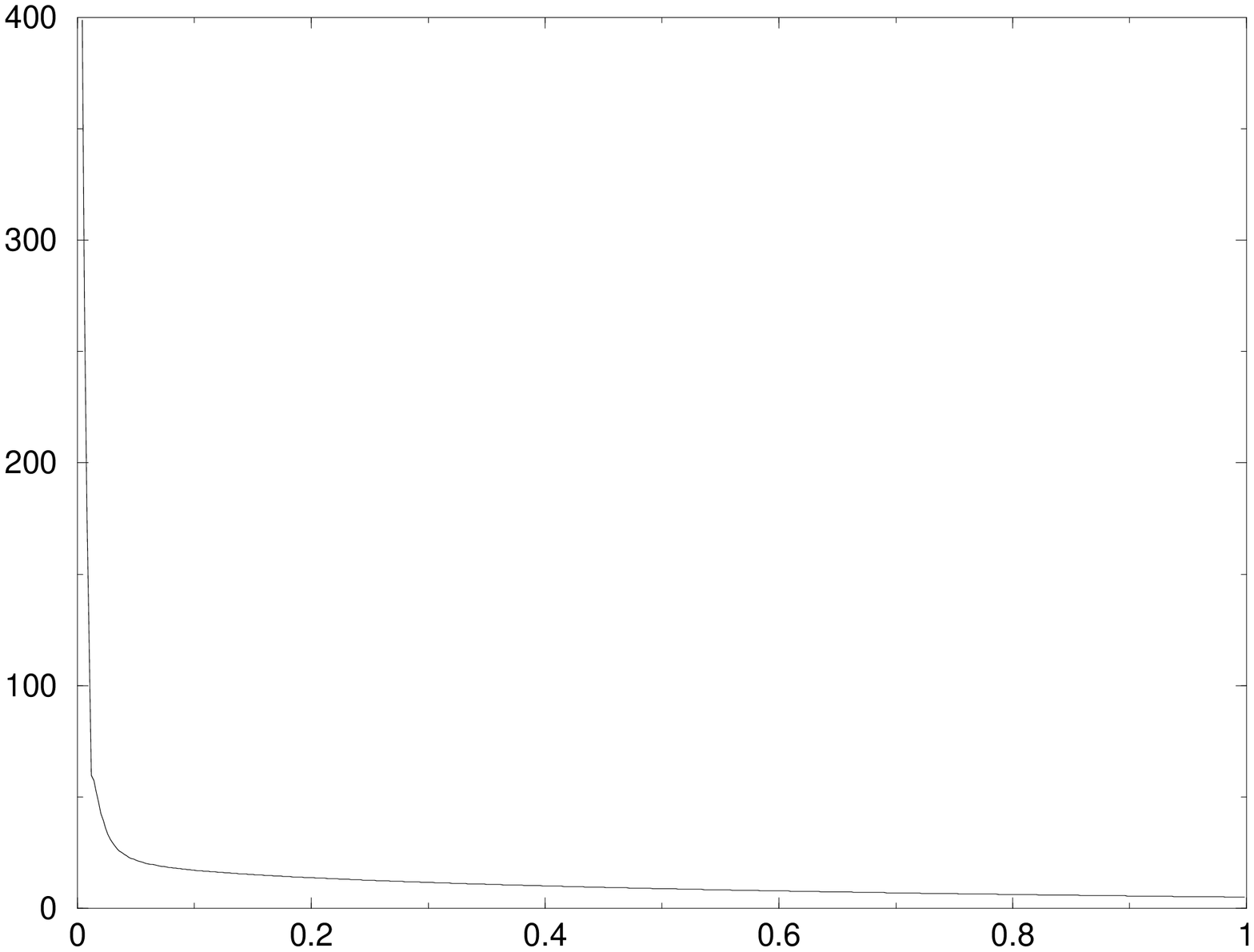}}\nonumber
\end{eqnarray}   
\vskip .5cm
\caption{{ Behaviour of $L_1$ with $\epsilon$, first for a parameter
 $l=1$, second diagram for $l=5$. We chose the parameter $r_+=10$. }}
\end{center}
\end{figure}

\begin{figure}[htb]
\begin{center}
\leavevmode
\begin{eqnarray}
\epsfxsize= 5truecm\rotatebox{-90}{\epsfbox{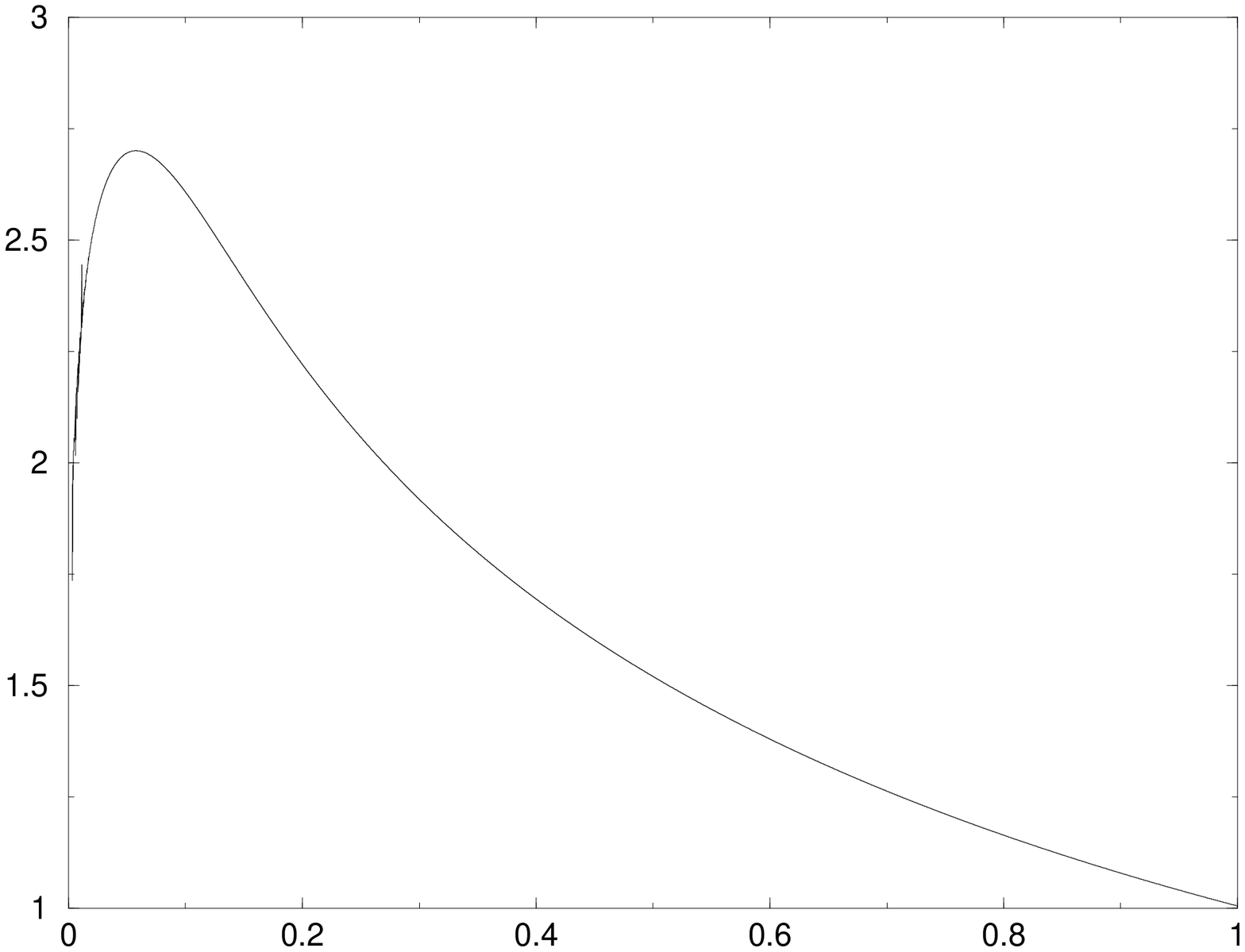}} & &
\epsfxsize=5truecm\rotatebox{-90}{\epsfbox{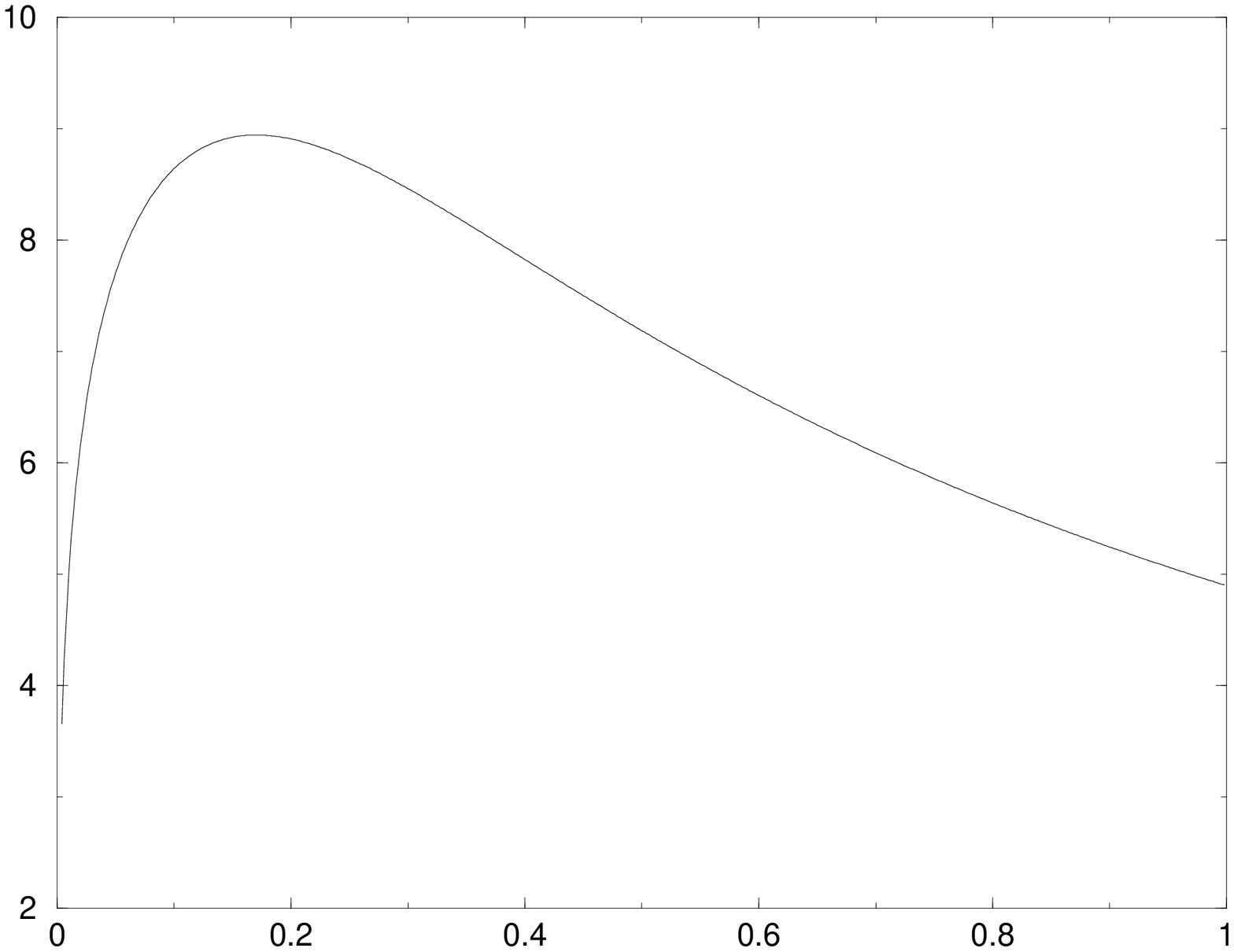}}\nonumber
\end{eqnarray}   
\vskip .5cm
\caption{{ Behaviour of $L_2$ with $\epsilon$, first for a parameter
 $l=1$, second diagram for $l=5$. We chose the parameter $r_+=10$. }}
\end{center}
\end{figure}

The characteristic of the proper distance can also be derived by studying
the limiting geometry of the extreme RN AdS black holes. The Euclidean
black hole metric can be written as 
\beq   
{\rm d}s^2=h{\rm d}\tau^2+h^{-1}{\rm d}r^2+r^2{\rm d}\Omega^2
\eeq
The Euclidean time takes its value in the range $0\leq\tau\leq T_H ^{-1}$.
Introducing a new variable $\tau_1=2\pi T_H\tau$ and $0\leq\tau_1\leq
2\pi$, we have 
\beq  
{\rm d}s^2=(\beta/2\pi)^2{\rm d}\tau_1^2+{\rm d}l^2+r^2{\rm d}\Omega^2
\eeq
where $\beta[r(l)]=\beta_H [h(r)]^{1/2}$ is the inverse local temperature
at an arbitrary point $r_+\leq r\leq r_B$. $l$ is the proper distance
between $r_+$ and $r$. We choose the coordinate according to 
\beq    
r-r_+=4\pi T_H b^{-1}\sinh^2(x/2), \hspace{0.5cm} b=h''(r_+)/2.
\eeq
In the limit $r_+\rightarrow r_B$, where the hole tends to occupy the
entire cavity, the region $r_+\leq r\leq r_B$ shrinks and we can expand
$h(r)$ in a power series $r-r_+$. After substituting Eqs.(7,4) into (11)
and taking the extreme limit in the end, we obtain
\beq      
{\rm d}s^2=r_B^2[\dpf{1}{1+6r_B ^2/l^2}({\rm d}\tau_1^2\sinh^2x+{\rm
d}x^2)+{\rm d}\Omega^2]
\eeq
In the Lorenzian version Eq(13) can be expressed as
\beq   
{\rm d}s^2=r_B^2[\dpf{1}{1+6r_B^2/l^2}(-{\rm d}t^2\sinh^2x+{\rm
d}x^2)+{\rm d}\Omega^2]
\eeq
This is the extension of the Bertotti-Robinson spacetime [22]. Taking
$\Lambda=-3/l^2\rightarrow 0$, it returns to the RN result in [4]. 

Now we are in a position to discuss the properties of the metric (14). In
the
extreme case the horizon is at $x=0$, as can be seen by noting that the
metric in (14) degenerates at $x=0$. The proper radial distance between
the horizon and any other point is finite, which is in agreement with the
numerical result exhibited in Fig.2.

Now we turn to concentrate our attention on the original RN AdS EBH. This
black hole satisfies Eq(6) at the very beginning. We put it in a cavity
with the boundary $r_B$. The expression of $h(r)$ now is
\beq       
h(r)=(1-r_+/r)^2+(r^2-4r_+^3/r+3r_+^4/r^2)/l^2
\eeq
Expanding the metric coefficients near $r=r_+$ and introducing
$r-r_+=r_B\rho^{-1}$ [4,8], we obtain
\beq  
{\rm d}s^2=r_B^2\rho^{-2}[-(1+6r_B^2/l^2){\rm d}t_1^2+{\rm
d}\rho^2+\rho^2{\rm d}\Omega^2]
\eeq
in the limit $r_+\rightarrow r_B$. Considering the vanishing cosmological
constant, it again boils down to (8) in the paper [4] for RN original EBH
case.

Using $r_B^2\rho^{-2}=0$ to determine the horizon, we find that the
horizon is infinitely far away ($\rho=\infty$). Therefore, the proper
distance
between the horizon and any other point at finite $\rho$ is infinite,
which agrees
to what is shown in Fig.1.

Different behaviors of the proper distances directly relate to different
Euler characteristics corresponding to the topology for RN AdS EBHs. The
Euclidean metric can be rewritten as 
\beq        
{\rm d}s^2=e^{2u(r)}{\rm d}\tau^2+e^{-2u(r)}{\rm d}r^2+R^2{\rm d}\Omega^2.
\eeq   
>From the Gauss-Bonnet (GB) theorem and the boundary condition, the Euler
characteristic $\chi$ takes the form [23,7]:
\beq   
\chi=\dpf{\beta_H}{2\pi}[(2u'e^{2u})(1-e^{2u}{R'}^2)]_{r_+}^{r_0}
\eeq
where $\beta_H=4\pi[(e^{2u})'_{r=r_+}]^{-1}$, and $e^{2u(r)}=h(r),
R^2=r^2$ in our case.

For Zaslavskii's treatment to obtain EBH, namely, by first taking the
boundary
limit and then extreme limit, we have
\beqn      
[(e^{2u}{R'}^2)\mid_{r=r_+=r_B}]_{extr} & = & 
[1-r_+/r-r_+^3/(l^2r)-Q^2/(r_+
r)+Q^2/r^2+r^2/l^2]_{r=r_+=r_B}\mid_{extr}  \no \\
& = & [1-r_B/r_B-r_B^3/(l^2 r_B)-Q^2/r_B^2+Q^2/r_B^2+r_B^2/l^2]_{extr}=0.
\eeqn
Since the horizon locates at a finite position, $\beta_H$ can be fixed,
therefore the Euler characteristic
$\chi=\dpf{4\pi}{2\pi}(\dpf{h'}{h'})_{r=r_+=r_B}\mid_{extr}=2$. This is
the same result as that of the NEBH.

However, for Hawking's treatment, starting with the original EBH, since
the horizon is infinitely far away, $\beta_H$ is not fixed. Under such a
condition we could get any value from the volume integral in the GB action
supplemented by the outer boundary contribution. As done for 4D RN black
hole as well as 4D charged dilaton black hole [23], the only way to get
a unique result is to add the inner boundary $r_0=r_+ +\ep$ and set
$\ep\rightarrow 0$ in the end of calculation. This will lead unambiguously
to $\chi=0$, which differs drastically from that of the NEBH result.

By means of the relation between the Euler characteristic and the entropy
first derived for NEBH [24] and later applied to EBHs [7,8]
\beq   
S=\dpf{A}{8}\chi
\eeq
and the different Euler characteristics obtained above for two different
treatments, natually we can naturally conclude that the RN AdS EBH with
nonextreme
topology has the entropy of $A/4$, while for the RN AdS EBH with extreme
topology, zero entropy emerges. These results can be used to explain the
intrinsic thermodynamical results obtained in [14-16].

In summary, we have shown that in the grand canonical ensemble, the 4D RN
AdS black hole can approach  the extreme state at nonzero temperature. The
proper distance and geometrical properties of the EBH developed from its
NEBH counterpart and the original EBH have also been exhibited. From
limiting metrics and Euler characteristics, we found that these two EBHs
are in different topological sectors, nonextreme and extreme
configurations, respectively. Due to different spacetime topology of these
two kinds of EBHs, it is easy to understand different intrinsic
thermodynamical properties, $S=A/4$ for EBH of NEBH topology and $S=0$ for
EBH of EBH topology. These results also affect the understanding of the
phase transition of RN AdS black holes. It has been shown that a phase
transition exists for a lot of black holes at the extreme limit [11-13].
Recently some classical critical phenomena has also been uncovered for RN
AdS black holes [18-19]. It would be reasonable to understand that phase
transition can happen between the RN AdS NEBH and EBH with a nonextreme
topological configuration. The RN AdS EBH with extreme topological
configuration is a quite different object as compared to the NEBH and thus
a phase transition cannot happen between them.

 ACKNOWLEDGEMENT: This work was partically supported
by Fundac\~ao de Amparo \`{a} Pesquisa do Estado de
S\~{a}o Paulo (FAPESP) and Conselho Nacional de Desenvolvimento 
Cient\'{\i}fico e Tecnol\'{o}gico (CNPQ).  B. Wang would also
like to acknowledge the support given by Shanghai Science and Technology
Commission.

\newpage

\end{document}